\documentclass[12pt]{article}
\usepackage{amssymb}
\newcommand{\rf}[1]{(\ref{#1})}
\newcommand{\bfa}{{\bf a}}\newcommand{\bb}{{\bf b}}\newcommand{\bc}{{\bf c}}
\newcommand{\bd}{{\bf d}}\newcommand{\be}{{\bf e}}\newcommand{\bm}{{\bf m}}\newcommand{\bn}{{\bf n}}
\newcommand{\bp}{{\bf p}}\newcommand{\bq}{{\bf q}}\newcommand{\br}{{\bf r}}\newcommand{\bv}{{\bf v}}\newcommand{\bu}{{\bf u}}\newcommand{\bx}{{\bf x}}

\newcommand{\bD}{{\bf D}}

\newcommand{\bI}{{\bf I}}\newcommand{\bM}{{\bf M}}\newcommand{\bP}{{\bf P}}\newcommand{\bQ}{{\bf Q}}
\newcommand{\bN}{{\bf q}}
\newcommand{\Dev}{{\mathbb D}ev}\newcommand{\E}{{\mathbb E}}
\newcommand{\bosy}[1]{ \mbox{\boldmath ${#1}$} }
\newcommand{\callig}{\mathbb}
\begin{document} 


\title{ 
\vspace{-0.8in}
\bf Acoustic axes in elasticity}
\author{Andrew N. Norris 
\\
Rutgers University,  Dept. of Mechanical \& Aerospace Engineering, \\ 98 Brett Road, Piscataway, NJ 08854-8058\, \, {\it norris@rutgers.edu} }
\date{}
\maketitle
\begin{abstract}
New results are presented for the degeneracy condition of elastic waves in anisotropic materials.  The condition for the existence of acoustic axes involves a traceless symmetric third order tensor that must vanish identically.  It is shown that all previous representations of the degeneracy condition follow from this {\it acoustic axis tensor}.   The  conditions for  existence of acoustic axes in  
 elastic crystals of orthorhombic, tetragonal, hexagonal and cubic (RTHC) symmetry are reinterpreted using the geometrical methods developed here.   
Application to weakly anisotropic solids is discussed, and it is shown that the satisfaction of the acoustic axes conditions to first order in anisotropy does not in general coincide with true acoustic axes.  

\end{abstract}

\section{Introduction}

 An acoustic axis is defined as a direction in which two sheets of the slowness surface intersect.  Propagation directions for which this occurs permit a pair of degenerate polarization vectors associated with the common slowness, and hence circularly polarized waves may propagate.  The importance of acoustic axes in crystals has been long appreciated, and the conditions required for their existence has been known since the early 1960's.  In a  paper on this subject in 1962, Khatkevich \cite{khatkevich1962} derived conditions for the existence of acoustic axes and discussed their application to the different crystal symmetry classes.  Acoustic axes are ``special directions" in the sense of Fedorov \cite{fed}, similar to directions in which transversely polarized waves propagate.  In fact, an acoustic axis is a direction for transverse waves, but not {\it vice versa} in general.    There has been renewed interest in acoustic axes in the past decade, initiated in part by Fedorov and Fedorov \cite{fedfed} who derived new and simplified representations of the acoustic tensor.   Based on this approach,  
Boulanger and Hayes \cite{bh98a,bh98} provided a complete classification of the acoustic axes in elastic crystals of orthorhombic, tetragonal, hexagonal and cubic (RTHC) symmetry.  In an interesting but unrelated paper, Mozhaev et al. \cite{mozhaev2001} gave an explicit condition for crystals of trigonal symmetry, thereby completing the classification for all crystal symmetry classes higher than monoclinic. 

The purpose here is not to attempt to extend this classification, but to focus on the general conditions and how they are formulated.  In particular we provide a definitive set of conditions which must be satisfied for acoustic axes to exist in an elastic material.  

A note on notation: lower case boldfaced latin symbols denote vectors, and vectors are normally of unit length, such as the unit direction vector $\bn$, and  the orthonormal triad $\{ \be_1, \be_2, \be_3\}$ which define the fixed directions.  

There are several ways of representing a a second order symmetric tensor $\bosy{\Lambda}$.  We begin with the perhaps less familiar Hamilton's cyclic form \cite{joly05}, also known as Hamilton's decomposition \cite{bhbiv} and the 
biaxial form \cite{fed}: 
\begin{equation} \label{s1}
\bosy{\Lambda} = \lambda \bI + \frac{\lambda '}{2} \big( \bfa_1 \otimes \bfa_2 + \bfa_2 \otimes \bfa_1 \big).  
\end{equation}
The numbers $\lambda$,  $\lambda '$, and the biaxial unit vectors $\bfa_1, \bfa_2$  are uniquely defined by $\bosy{\Lambda}$.  The proof of eq. \rf{s1} may be seen by direct construction, see Appendix A.  Fedorov \cite{fed} calls the directions $\bfa_1$ and $\bfa_2$ the axes of the tensor (not to be confused with the principal axes).  
The standard diagonal form for $\bosy{\Lambda}$, also known as the Gibbs decomposition \cite{bhbiv}, employs the principal axes, an orthonormal triad $\{\bfa, \bb , \bc\}$, 
\begin{equation} \label{ss3}
\bosy{\Lambda} =  \lambda_1\, \bfa\otimes \bfa + \lambda_2\, \bb\otimes \bb + \lambda_3\, \bc\otimes \bc,
\end{equation}
with the eigenvalues ordered $\lambda_1 \ge \lambda_2 \ge \lambda_3 $.  Equations \rf{s1} and \rf{ss3} can be reconciled  
using the identity $\bfa\otimes \bfa +   \bb\otimes \bb +   \bc\otimes \bc = \bI$, implying  $\lambda=\lambda_2$,  $\lambda '=\lambda_1- \lambda_3$, and 
\begin{equation} \label{s6}
\bfa_j = \big(\frac{\lambda_1-\lambda_2}{\lambda_1-\lambda_3}\big)^{1/2}\ \bfa + (-1)^j \big(\frac{\lambda_2-\lambda_3}{\lambda_1-\lambda_3}\big)^{1/2}\ \bc\, , \quad j=1,2.
\end{equation}

Any positive definite symmetric tensor $\bosy{\Lambda}$ defines an ellipsoidal surface $\{\bx : \bx \cdot \bosy{\Lambda} \bx = 1\}$.  
A symmetric second order tensor is   $uniaxial$ if two of its eigenvalues coincide and the third corresponds to the preferred axial direction.  The ellipsoid reduces to a spheroid, which may be oblate or prolate, depending as $
\lambda_1 = \lambda_2 > \lambda_3$ (oblate) or $\lambda_1 > \lambda_2 = \lambda_3$ (prolate).  In either event,  $\bosy{\Lambda}$ reduces to 
\begin{equation} \label{s7}
\bosy{\Lambda} = \lambda_2 \bI \pm \lambda ' \bd \otimes  \bd , \qquad \mbox{uniaxial}, 
\end{equation}
where, according to eq. \rf{s1}, the $\pm$ corresponds to $\bfa_1 = \pm \bfa_2$, and 
the implications for the eigenvalues $\lambda_1$, $\lambda_2$, and $\lambda_3$ and the definition of $\bd$ follow from eq. \rf{s6}. 
Note that the sign of the $\bd \otimes  \bd$ term in eq. \rf{s7} directly indicates whether the uniaxial tensor is prolate $(+1)$ or oblate $(-1)$.   In the limit as the three eigenvalues coincide 
 eqs. \rf{s1}, \rf{s6} and \rf{s7} all reduce to $\bosy{\Lambda} = \lambda_2 \bI$ as $\lambda ' \rightarrow 0$, i.e. $\bosy{\Lambda}$ is  spherical.   An oblate (prolate) acoustical tensor corresponds to one where the two larger (lesser) phase speeds coincide for a given choice of the phase direction.  It is  worth mentioning that the geometric descriptions (oblate, etc.) do not refer to the slowness surface but  to the form of the acoustical tensor for the given phase direction.

Our focus will be on discussing the conditions under which the acoustic tensor of elasticity, $\bQ$ defined below, is uniaxial.   Before doing so, it is useful to note that it is relatively simple to define conditions for sphericity.  Thus,  the identity 
\begin{equation} \label{egg}
(\lambda_1 - \lambda_2)^2 + (\lambda_2 - \lambda_3)^2 +(\lambda_3 - \lambda_1)^2  = 2\, I^2 - 6\, II\, ,  
\end{equation}
where $I$ and $II$ are invariants of  $\bosy{\Lambda}$, $ I = \mbox{tr }\bosy{\Lambda}$, $2 II =  (\mbox{tr}\bosy{\Lambda})^2 -  \mbox{tr}\bosy{\Lambda}^2$, 
implies that  all three eigenvalues coincide iff the condition  $I^2=3II$ is met.  Quantities such as 
$(\lambda_1 - \lambda_2)^3 + (\lambda_2 - \lambda_3)^3 +(\lambda_3 - \lambda_1)^3 $ 
and $\Gamma = (\lambda_1 - \lambda_2)(\lambda_2 - \lambda_3)(\lambda_3 - \lambda_1)$  vanish iff   $\bosy{\Lambda}$ is uniaxial.  However, it does not seem possible to represent these in terms of the invariants of a second order tensor: $I$, $II$ and the third scalar invariant  $III = \mbox{det}\bosy{\Lambda}$.  We will find instead that a third order tensor is required.

\section{Acoustic axes }\label{secaa}
The acoustical tensor of elasticity,  $\bQ (\bn)$,  is symmetric positive definite and defined as 
$ Q_{ik} (\bn) = C_{ijkl}n_jn_l$, where $\bf C$ with elements $C_{ijkl}$ is the elastic stiffness. The eigenvalues of $\bQ$ are $\lambda_j = \rho v_j^2$, where $\rho $ is the mass density and $v_j (\bn)$, $j=1,2,3$ are the   phase speeds for phase direction $\bn$.    
According to the general representation of eq. \rf{s7}, at an acoustic axis, $\bQ$ must be of the form
\begin{eqnarray} \label{form1}
\left\{ \begin{array}{c}
Q_{11}\\ Q_{22}\\ Q_{33}\\ Q_{23} \\ Q_{31} \\Q_{12}
\end{array} \right\} = 
\lambda_2 \, 
\left\{ \begin{array}{c}
1 \\ 1 \\ 1\\ 0 \\ 0 \\ 0 
\end{array} \right\}
\pm
\left\{ \begin{array}{c}
q_1^2 \\  q_2^2 \\ q_3^2 \\ q_2 q_3 \\ q_3 q_1 \\ q_1 q_2
\end{array} \right\} , 
\end{eqnarray}
where we have chosen for purposes of clarity to write the six elements of $\bQ$ as a column ``vector".   
The form of $\bQ$ in eq. \rf{form1} is motivated by the exhaustive study of Boulanger and Hayes \cite{bh98a,bh98} for RTHC crystals.  Boulanger and Hayes, using general expressions derived by Fedorov and Fedorov \cite{fedfed}, wrote $\bQ$ as in eq. \rf{form1} and  obtained explicit results for the number and orientation of acoustic axes in RTHC crystals.    We do not restrict the present analysis to any particular elastic symmetry, although we will return to the RTHC crystals later in Sections \ref{spec} and \ref{gi}. 
Focusing on the final three elements in $\bQ$, implies that $q_2 q_3 = Q_{23}$, etc.   For the moment let us assume that all three of $Q_{23},\, Q_{31},\, Q_{12}$ are non-zero.  Define the vector $\bq (\bn) = q_1\be_1+q_2\be_2+ q_3\be_3$ by 
\begin{eqnarray} \label{by}
q_1 &=& \big( s\, \frac{Q_{31}Q_{12}}{Q_{23}}\big)^{1/2}\, \mbox{sign}\, Q_{23}, 
\nonumber \\
q_2 &=& \big( s\, \frac{Q_{12}Q_{23}}{Q_{31}}\big)^{1/2}\, \mbox{sign}\, Q_{31}, 
\\
q_3 &=& \big( s\, \frac{Q_{23}Q_{31}}{Q_{12}}\big)^{1/2}\, \mbox{sign}\, Q_{12}, 
\nonumber  
\end{eqnarray} 
where $s = \mbox{sign}\, Q_{23}Q_{31}Q_{12}$. 
The  acoustic tensor then can be written,  
\begin{equation} \label{w2}
\bQ (\bn) = \bD + s\, \bq\otimes \bq ,  
\end{equation}
where $\bD$ is diagonal with $D_i = Q_{ii} - s\, q_i^2$ (no sum), that is, 
\begin{equation} \label{w4} 
D_1 = Q_{11} -\frac{ Q_{31}Q_{12}}{Q_{23}}, \quad 
D_2 = Q_{22} - \frac{Q_{12}Q_{23}}{Q_{31}}, \quad
D_3 = Q_{33} - \frac{Q_{23}Q_{31}}{Q_{12}}. 
\end{equation}
The expression of the acoustical tensor in the form \rf{w2}, while entirely general only as long as $Q_{23}Q_{31}Q_{12}\ne 0$,  is very useful for examining at this stage some, but not all, of the conditions under which acoustic axes occur.   

In particular,   $\bQ$ is uniaxial when $\bD$ is spherical, that is: $D_1=D_2=D_3$. If this occurs, 
then the uniaxial acoustical tensor is prolate or oblate depending as $s=+1$ or $s=-1$, respectively.  This implies that the two outer (inner) sheets of the slowness surface coalesce if $s=+1$ ($s=-1$).  Furthermore, the vector $\bq$ is in the direction of the third polarization, perpendicular to the plane of polarization of circularly polarized waves. 
Equating the $D_i$ implies the  conditions,  
\begin{eqnarray} \label{w6b}
D_2=D_3 \quad \Rightarrow\quad  Q_{22} - Q_{33}&=&  Q_{23}\, \big( \frac{Q_{12}}{Q_{31}} - \frac{Q_{31}}{Q_{12}} \big) ,
\nonumber \\
D_3=D_1 \quad \Rightarrow\quad Q_{33} - Q_{11}&=&  Q_{31}\, \big( \frac{Q_{23}}{Q_{12}} - \frac{Q_{12}}{Q_{23}} \big)  ,
\\
D_1=D_2 \quad \Rightarrow\quad Q_{11} - Q_{22}&=&  Q_{12}\, \big( \frac{Q_{31}}{Q_{23}} - \frac{Q_{23}}{Q_{31}} \big)  .
\nonumber 
\end{eqnarray}
In order to account for the possibility that  the denominator vanishes, these equations are normally expressed \cite{khatkevich1962}
\begin{equation} \label{w6}
\phi_j= 0, \quad j=1,2,3, 
\end{equation}
where $\phi_j$ are found by multiplying eqs. \rf{w6b} by  $Q_{12}Q_{23}Q_{31}$, 
\begin{eqnarray} \label{w6a}
\phi_1 &=& Q_{31}Q_{12}\, \big( Q_{33} - Q_{22}\big) + Q_{23}\, \big( Q_{12}^2 - Q_{31}^2\big) , 
\nonumber \\
\phi_2 &=&Q_{12}Q_{23}\, \big( Q_{11} - Q_{33}\big) + Q_{31}\, \big( Q_{23}^2 - Q_{12}^2\big) , 
\\
\phi_3 &=&Q_{23}Q_{31}\, \big( Q_{22} - Q_{11}\big) + Q_{12}\, \big( Q_{31}^2 - Q_{23}^2\big) .
\nonumber 
\end{eqnarray}
Equations \rf{w6} were first derived by Khatkevich in 1962 \cite{khatkevich1962} by a different method.     Khatkevich argued that the orientation of acoustic axes in general follows from the condition that the cofactor matrix of  
$\bM = \bQ - \lambda \bI$ must vanish.  Only three of the elements of the cofactor matrix are independent because of the simultaneous requirement that det$\, \bM = 0$.  Setting the off-diagonal elements of the cofactor matrix to zero implies \cite{mozhaev2001}
\begin{equation} \label{m1}
\left| \begin{array}{cc}
Q_{11} - \lambda & Q_{12}\\
Q_{13} & Q_{23} \end{array} \right | = 
\left| \begin{array}{cc}
Q_{12}  & Q_{22}- \lambda \\
Q_{13} & Q_{23} \end{array} \right | =
\left| \begin{array}{cc}
Q_{12} & Q_{23}\\
Q_{13} & Q_{33}- \lambda  \end{array} \right | =0\, , 
\end{equation}
and eliminating $\lambda$ from these recovers Khatkevich's conditions \rf{w6}.  There has been some confusion about the starting point for Khatkevich's derivation, i.e., the fact that cof$\, \bM = 0$ at an acoustic axis.  Mozhaev et al. \cite{mozhaev2001} note that this assumption is not proved explicitly by Khatkevich or in the references that he offered, and they suggest that the arguments given by Fedorov for this condition  are ``rather tangled".  However, at an acoustic axis $\bQ$ has the form of \rf{w2} with $\bD = \lambda \bI$. Therefore $\bM  = s\, \bq\otimes \bq$ is of rank one, and the cofactor matrix is identically zero: cof\, $M_{ij} = e_{ikl}e_{jmn} q_k q_l q_m q_n = 0$. 

It has been argued \cite{khatkevich1962} that  only two of the three conditions \rf{w6} are independent, since, for example, the third equality $D_1=D_2$ follows from the previous two in eqs. \rf{w6a}.  However, the  conditions \rf{w6} are not always sufficient, as for instance, on a plane of symmetry.  Consider $\bn$ lying in a material plane of symmetry: to be specific, suppose $\be_3$ is normal to a plane of material symmetry and $n_3=0$.  Then $Q_{23}$ and $Q_{31}$ both vanish, and the three conditions \rf{w6} are identically satisfied.   An additional equation is clearly necessary.  One may be constructed by taking the product of the first two identities in eq. \rf{w6b}: 
\begin{equation} \label{prod}
\big( Q_{11} - Q_{33}\big) \big( Q_{22} - Q_{33}\big)=  Q_{12}^2\, \big( 1 - \frac{Q_{23}^2}{Q_{12}^2}  \big) \big( 1 - \frac{Q_{31}^2}{Q_{12}^2}  \big)   .
\end{equation}
If the elastic body possesses a plane of symmetry with normal $\bf e$, and if $\bn$ lies in the plane of symmetry, then ${\bf Q}({\bf n})$ must display the same symmetry, viz. $\bm\cdot{\bf Qe}=0$ for all $\bm\perp\be$.   With ${\bf e} = \be_3$, and $n_3=0$, eq. \rf{prod}  becomes
\begin{equation} \label{proda}
\big( Q_{11} - Q_{33}\big) \big( Q_{22} - Q_{33}\big)-   Q_{12}^2 = 0  .
\end{equation} 
This condition was also derived by Khatkevich \cite{khatkevich1962}, using a different argument.  We show later that it is a consequence of another condition (see eq. \rf{qex}), and will return to it after a more consistent and thorough set of conditions have been discussed.  The point at this stage is to emphasize the multiplicity of conditions, some for general directions, others for planes of symmetry.  This apparent complexity raises the question of what are the minimum number of conditions, and how may they be expressed?   One solution was offered by Al'shits and Lothe \cite{alshits1979a} who derived a set of conditions in terms of a seven component vector  which must vanish for acoustic axes to occur in a given direction $\bn$. An alternative solution to the question will be provided next, and  the Al'shits and Lothe conditions will be considered later and  shown to be overly restrictive. 

\section{A coordinate invariant set of conditions for acoustic axes}\label{cis}

Consider a symmetric tensor $\bosy{\Lambda}$   with orthonormal eigenvectors $\bfa$, $\bb$ and $\bc$, and  distinct  eigenvalues  $\lambda_i,\, i=1,2,3$.  Noting that   
\begin{equation} \label{ac2}
\bosy{\Lambda}^2 = \lambda_1^2\, \bfa\otimes \bfa +\lambda_2^2\, \bb\otimes \bb +\lambda_3^2\, \bc\otimes \bc \, ,
\end{equation}
implies, for instance,  that   $\bosy{\Lambda}\bm$ and $ \bosy{\Lambda}^2\bm \in $ span$\{\bfa , \bb\}$ for any  vector $\bm$  perpendicular to $\bc$.  
The triple product of three vectors,  $[\bp , \, \bq ,\, \br ] = e_{ijk}p_iq_jr_k$, is zero if the vectors are coplanar.  Hence, if $\bm \in $ span$\{\bfa , \bb\}$,  it  follows that $[ \bm,\, \bosy{\Lambda}\bm,\, \bosy{\Lambda}^2\bm ]$ is zero.  Consequently,  the vanishing of 
\begin{equation} \label{deff}
 [ \bm,\, \bosy{\Lambda}\bm,\, \bosy{\Lambda}^2\bm ]  
\end{equation}
 is a necessary consequence of the fact that $\bm$ is perpendicular to one of the eigenvectors of $\bosy{\Lambda}$.  It remains to show that it is also a sufficient condition. 

In order to understand the function of $\bm$ in eq. \rf{deff}   first consider the more general quantity $[ \bm,\, \bosy{\Lambda}\bp,\, \bosy{\Lambda}^2\bq ] = \bm\cdot \big( \bosy{\Lambda}\bp\wedge \bosy{\Lambda}^2\bq\big)$ for arbitrary $\bm$, $\bp$ and $\bq$.    
Direct calculation based on  eq. \rf{ac2}, and the fact that $\{\bfa , \, \bb ,\, \bc \}$
 form an orthonormal triad, gives
\begin{eqnarray} \label{ac4}
\bosy{\Lambda}\bp\wedge \bosy{\Lambda}^2\bq 
&=& \big(  \lambda_3\, \bb\cdot\bp\, \bc\cdot\bq - \lambda_2\, \bc\cdot\bp\, \bb\cdot\bq
\big)\, \lambda_2\lambda_3\bfa +
\nonumber \\&&  \big
(  \lambda_1\, \bc\cdot\bp\, \bfa\cdot\bq - \lambda_3\, \bfa\cdot\bp\, \bc\cdot\bq
\big)\, \lambda_3\lambda_1\bb  +
\nonumber \\
&&  \big(  \lambda_2\, \bfa\cdot\bp\, \bb\cdot\bq - \lambda_1\, \bb\cdot\bp\, \bfa\cdot\bq
\big)\, \lambda_1\lambda_2\bc\, . 
\end{eqnarray}
The general form of \rf{deff} follows from the identity \rf{ac4} as 
\begin{equation} \label{ac5}
[ \bm,\, \bosy{\Lambda}\bm,\, \bosy{\Lambda}^2\bm ]
= (\lambda_1-\lambda_2) (\lambda_2 - \lambda_3) (\lambda_3 - \lambda_1 ) \, 
(\bm\cdot\bfa)\, (\bm\cdot\bb) \,(\bm\cdot\bc)\, . 
\end{equation} 
By assumption, the eigenvalues are distinct, implying that the factor premultiplying $(\bm\cdot\bfa)\, (\bm\cdot\bb) \,(\bm\cdot\bc)$ is non-zero.   Therefore, if the triple product in eq. \rf{ac5} vanishes the vector $\bm$ must be perpendicular to at least one of $\bfa$, $\bb$ or $\bc$.  This proves that the triple product \rf{deff} vanishes iff $\bm$ is orthogonal to one of the eigenvectors of $\bosy{\Lambda} $ when the eigenvalues are distinct.   

Now consider the special case that  $\bosy{\Lambda}$ is uniaxial, say $\lambda_2 = \lambda_3$. Then every direction perpendicular to $\bfa$ is an eigenvector:  this is circular polarization corresponding to the circle of eigenvectors $\{\bp : \, |\bp|=1,\, \bp\cdot\bfa=0\}$.  Conversely, for a given arbitrary direction $\bm$, there is an eigenvector on this circle that is perpendicular to $\bm$, specifically $\bp \parallel \bm\wedge\bfa$.   Hence, 
$[ \bm,\, \bosy{\Lambda}\bm,\, \bosy{\Lambda}^2\bm ] $ vanishes for any and all directions $\bm$ if 
$\bosy{\Lambda} $ is uniaxial.  
The converse follows from the explicit form \rf{ac5}.  Thus, if the triple product vanishes for all $\bm$ then 
\begin{equation} \label{ac7}
(\lambda_1-\lambda_2) (\lambda_2 - \lambda_3) (\lambda_3 - \lambda_1 ) = 0 , 
\end{equation}
which implies that at least two of the eigenvalues are equal.  This proves the fundamental result:

{
\parbox{\columnwidth}{
{\bf Uniaxial Condition}: The symmetric tensor $\bosy{\Lambda}$ is uniaxial if and only if the triple product $[ \bm,\, \bosy{\Lambda}\bm,\, \bosy{\Lambda}^2\bm ]$ vanishes for all $\bm$.} }

This can be immediately applied to the acoustic tensor $\bQ (\bn)$, yielding      the following criterion for the existence of an acoustic axis: 
\begin{equation} \label{ca1}
\bn \mbox{ is an acoustic axis} \quad \iff \quad f(\bm , \bQ (\bn) ) = 0 , \ \  \forall \ \bm ,
\end{equation}
where  the function $f$ is defined 
\begin{equation} \label{ac8}
f(\bm ) = f(\bm , \bQ (\bn) ) = [ \bm,\, \bQ \bm,\, \bQ ^2\bm ] . 
\end{equation}
Referring to eq. \rf{ac5} we see that this is in general an overly stringent condition.  Specifically, it need not apply to all directions $\bm$ but only to a sufficient number of directions   chosen so that $(\bm\cdot\bfa)\, (\bm\cdot\bb) \,(\bm\cdot\bc) \ne 0$.   This requires that $\bm$ not be limited to the set $\{\bfa,\, \bb ,\, \bc\}$, which is satisfied if, for instance, eq. \rf{ca1} holds for  four $\bm$, as discussed in Section \ref{aat}.  

\section{The acoustic axis tensor}\label{aat}

The function $f(\bm , \bQ (\bn) )$ is third order in $\bm$ and defines a third order tensor $\bosy{\phi}(\bn)$ via eqs. \rf{ac5} and \rf{ac8}:
\begin{equation} \label{phi}
\bosy{\phi} = (\lambda_1-\lambda_2) (\lambda_2 - \lambda_3) (\lambda_3 - \lambda_1 )\, 
\big\{\bfa\bb\bc\big\}\, ,
\end{equation}
where $\big\{\bfa\bb\bc\big\}$ is the totally symmetric part \cite{backus} of $\bfa\otimes\bb\otimes\bc$, i.e. 
\begin{equation} \label{phi2}
\big\{\bfa\bb\bc\big\}_{ijk} = \frac16\, \big(
a_ib_jc_k+a_jb_kc_i+a_kb_ic_j+a_ib_kc_j+a_jb_ic_k+a_kb_jc_i\big). 
\end{equation}
In terms of the fixed coordinate system, 
\begin{equation} \label{ca4}
f(\bm) = \phi_{ijk}\, m_im_jm_k, \quad\mbox{and}\quad \bosy{\phi} = \phi_{ijk}\, \be_i\otimes \be_j\otimes \be_k \, . 
\end{equation}

The tensor $\bosy{\phi}$ is totally symmetric, that is, $\phi_{ijk}$ is unchanged under any permutation of the suffices $i,j,k$ 
and hence has 10 distinct elements $ \phi_{ijk}$.  However, from its definition in eq. \rf{phi}, 
\begin{equation} \label{phi3}
\phi_{ikk} = 0 \quad \mbox{for }i=1,2,3. 
\end{equation}
A totally symmetric tensor with this traceless property is called {\em harmonic} \cite{backus} by virtue of the fact that $\nabla^2 
\phi_{ijk}x_ix_jx_k = 0$,  which follows from eq. \rf{phi3}.  The three relations \rf{phi3} imply that $\bosy{\phi}$ is characterized by seven independent elements.  These may be expressed via an explicit decomposition of $\Dev$, the space of symmetric traceless third order tensor.  This is achieved by a partition $\Dev$ into a direct sum of spaces of one or two dimensions each of which is irreducible under rotation about an axis, i.e. the group SO(2).  The procedure, called ``Cartan decomposition"  when applied to harmonic polynomials,    has been found useful in describing the symmetry classes of  elasticity  \cite{FV} and photoelasticity  \cite{FV97}. 

Specifically, $\Dev$ is the direct sum of ${\callig{D}}_0$, ${\callig{D}}_1$, ${\callig{D}}_2$ and ${\callig{D}}_3$, the spaces spanned, respectively, by $\E_0$, and the pairs $(\E_1, \, \E_2 )$,  $(\E_3, \, \E_4 )$ and $(\E_5, \, \E_6 )$, where
\begin{eqnarray} \label{irr2}
&& \E_0 =  (10)^{-1/2}\,\big\{( 5\be_3^2 - 3\bI) \be_3 \big\},
\nonumber \\
&& \E_1 =(3/20)^{1/2}\, \big\{(5\be_3^2 - \bI )\be_1 \big\}, \qquad \E_2 =(3/20)^{1/2}\,\big\{ (5\be_3^2 - \bI )\be_2 \big\},
\nonumber \\
&& \E_3 =(3/2)^{1/2}\,\big\{( \be_1^2 - \be_2^2) \be_3 \big\}, \quad \E_4 =6^{1/2}\,\big\{\be_1 \be_2 \be_3 \big\},
 \\
&& \E_5 =\frac12\, \big\{\be_1^3 - 3\be_1\be_2^2 \big\}, \quad \E_6 =\frac12\, \big\{\be_2^3 - 3\be_1^2\be_2 \big\}.
\nonumber
\end{eqnarray}
  The seven spaces are grouped into the subspaces ${\callig{D}}_n$, $n=0,1,2,3$, because of the property that  elements of the associated spaces are rotated through angle $n\theta$ under a coordinate rotation of $\theta$ about the axis $\be_3$.  The reader is referred to Appendix B for a derivation of this result.   Furthermore, the tensors $\E_n$ form an orthonormal basis of $\Dev$, and hence the general representation of the acoustic axis tensor, or any harmonic third order tensor, is 
\begin{equation} \label{gen}
\bosy{\phi} = \sum\limits_{n=0}^6\, c_n\, \E_n \quad \mbox{where} \quad
c_n =  \E_n\cdot \bosy{\phi} \, . 
\end{equation}
Direct calculation using eqs. \rf{ca4} and \rf{irr2} yields
\begin{eqnarray} \label{gen1}
& c_0 = (5/2)^{1/2}\, \phi_{333} , 
& c_1 = (15/4)^{1/2}\,\phi_{133}, 
\nonumber \\
& c_2 = (15/4)^{1/2}\, \phi_{233},  
& c_3 = (3/2)^{1/2}\, (\phi_{113} - \phi_{223}),
\\
& c_4 = 6^{1/2}\, \phi_{123}, \quad \quad \quad & c_5 = (\phi_{111} - 3\phi_{122})/2, 
\nonumber \\
& c_6 =  ( \phi_{222} - 3\phi_{112})/2 .  & {  } \nonumber
\end{eqnarray}
Inner products of two tensors are $\bosy{\phi}\cdot\bosy{\phi}' \equiv \phi_{ijk}\phi_{ijk}' = c_0c_0'+ c_1c_1' + \ldots + c_7c_7'$.  The strict condition for acoustical degeneracy, $\bosy{\phi} = 0$ is therefore equivalent to the requirement that the seven coefficients $c_j$, $j=0,1,2,\ldots , 7$ vanish \cite{alshits1979a}.   This also follows from the general identity
\begin{equation} \label{idn}
  |\bosy{\phi}|^2 = \sum\limits_{n=0}^6
 c_n^2. 
 \end{equation}

The elements $\bosy{\phi} $ may be expressed in terms of the elements of the acoustical tensor $\bQ(\bn)$.   Direct calculation using eq. \rf{ac8} gives
\begin{equation} \label{q1}
\phi_{iii} = f(\be_i) = \phi_i \quad\mbox{(no sum)}, 
\end{equation}
where $\phi_i,\, i=1,2,3$, are defined in eq. \rf{w6a}.  Similarly, it can be shown that 
\begin{eqnarray} \label{q2}
3 \phi_{iij}  &=& \big[ (Q_{jj} - Q_{ii} ) \phi_i - Q_{ij} \phi_j + Q_{ik}\phi_k\big] / Q_{ij}\, ,\quad  i\ne j\ne k\ne i \quad \mbox{(no sum)}, 
\nonumber \\ && \\
6 \phi_{123} &=&(Q_{11}Q_{22} - Q_{12}^2)(Q_{22} - Q_{11})\, +\, (Q_{22}Q_{33} - Q_{23}^2)(Q_{33} - Q_{22})
\nonumber \\ && \, +\, 
(Q_{33}Q_{11} - Q_{31}^2)(Q_{11} - Q_{33})\, . 
\nonumber
\end{eqnarray}
For instance, $\phi_{123}$ is the only nonzero element of $\bosy{\phi}$ if $\bQ$ is diagonal.  Also, if $Q_{23}$ and $Q_{31}$ are zero, as is the case when the material has a plane of symmetry with normal $\be_3$ and $\bn\perp\be_3$, then 
\begin{equation} \label{qex}  \phi_{\alpha \alpha 3} = - (-1)^\alpha\, \big[ ( Q_{11} - Q_{33}) ( Q_{22} - Q_{33})-   Q_{12}^2 \big] Q_{12}/3 , 
\quad \alpha = 1 \mbox{ or } 2.  
\end{equation}
Hence, the condition \rf{proda}, which was derived in an {\it ad hoc} manner,  can be viewed as a consequence of $ \phi_{113} = 0$ or $ \phi_{223} = 0$. 

The elements $\phi_{ijk}$ are related to the seven quantities $R_i,\, i=1,2,\ldots , 7$ of Al'shits and Lothe \cite{alshits1979a} as follows: 
\begin{eqnarray} \label{equ}
&R_1 = -\phi_{333}, \quad
R_2 = \phi_{222}, \quad
R_3 = -\phi_{111}, \quad R_4 = -\phi_{111} - 3\phi_{133}, &
\nonumber \\
& 
R_5 = \phi_{222} + 3\phi_{233}, \quad
R_6 = -\phi_{333} - 3\phi_{223}, \quad R_7 =  -6\phi_{123}  .&
\end{eqnarray}
There is a typographical error in eq. $(29)_3$ of Al'shits and Lothe \cite{alshits1979a}: the term $(-3R_3-2R_4)$ should be $(-3R_3-R_4)$. 

We mention some other properties of the tensor $\bosy{\phi}$, beginning with 
\begin{equation} \label{am}
|\bosy{\phi}|^2 =  \frac16\, \Gamma^2 , \quad \mbox{where} \quad \Gamma = (\lambda_1-\lambda_2) (\lambda_2 - \lambda_3) (\lambda_3 - \lambda_1 ) \, ,  
\end{equation}
and the related result,  from eq. \rf{phi}, that 
\begin{equation} \label{nn}
|f(\bm)| \le \frac{\sqrt{2}}{3} \, |\bosy{\phi}| \, |\bm|^3 , 
\end{equation}
with equality iff $|\bfa\cdot \bm |= | \bb\cdot \bm |=| \bc\cdot \bm | = |\bm|/\sqrt{3}$. 
In addition to being traceless, eqs. \rf{phi} and \rf{phi2} imply that $\bosy{\phi}$ satisfies the orthogonality-like relations   
\begin{equation} \label{pro}
\phi_{ipq}\phi_{jpq} = \frac{1}{18} \Gamma^2\, \delta_{ij}. 
\end{equation}
Define the 6-vectors $\bv_i$, $i=1,2,3$, 
\begin{eqnarray} \label{6v}
\bv_i =   \left\{ 
\phi_{i11},\,  \phi_{i22},\,  \phi_{i33},\,  \sqrt{2}\phi_{i23},\,  \sqrt{2}\phi_{i31},\,  \sqrt{2}\phi_{i12}  
\right\}^T, \quad i=1,2,3. 
\end{eqnarray}
Then eq. \rf{pro} may be rewritten in a form that appears more like an orthogonality property: 
\begin{equation} \label{pro2}
\bv_i \cdot \bv_j =  \frac13  |\bosy{\phi}|^2\, \delta_{ij},\quad i,j=1,2,3,
\end{equation}
while the identities \rf{phi3} become
\begin{equation} \label{idb}
\bv_i \cdot \br =  0,\quad i=1,2,3, \quad \br=(1,\,1,\,1,\,0,\,0,\, 0)^T.  
\end{equation}
Using for instance, eqs. \rf{pro2} and \rf{idb} with $i=1$,  the magnitude of the acoustic axis tensor can be expressed by means of  no more than five of the ten components of $\phi_{ijk}$ (compare with the expression \rf{idn} that employed 7 coefficients):
\begin{equation} \label{pro3}
 \frac13\, |\bosy{\phi}|^2 = \phi_{111}^2+\phi_{122}^2+(\phi_{111}+ \phi_{122})^2+
 2\phi_{123}^2+ 2\phi_{112}^2+\phi_{113}^2\, . 
\end{equation}

The tensor $\bosy{\phi}(\bn)$ must vanish identically if $\bn$ is an acoustic axes.  In principle, this requires all seven components $c_n,\, n=0,1,2,\ldots , 6$ must be zero, or all five of the components in eq. \rf{pro3}.   In practice, in order to use the function $f(\bm)$ to test whether or not two eigenvalues coincide, the direction $\bm$ must be chosen so that the product $g(\bm) \equiv (\bfa\cdot \bm) (\bb\cdot \bm)(\bc\cdot \bm)$ is non-zero.  For an arbitrary choice of $\bm$ this could vanish if $\bm$ lies in one of the planes normal to $\bfa$, $\bb$ or $\bc$.  Given two or three  arbitrary directions, $g$ could vanish for all if they each lie in  these planes.  However, for four directions, such as $\be_1$, $\be_2$, $\be_3$, and $\be_1+\be_2+\be_3$, then at least one of them must yield a non-zero $g$. Note that $f(\be_i) = \phi_{iii}$ (no sum), and using  identities like 
\begin{equation} \label{eg2}
f(\be_1+\be_2 \pm \be_3) = 6\phi_{123} - 2( \phi_{111}+\phi_{222}\pm \phi_{333}), 
\end{equation} 
we surmise that the simultaneous vanishing of, e.g., $\phi_{111}$, $\phi_{222}$, $\phi_{333}$ and $\phi_{123}$, is sufficient to ensure the existence of an acoustic axis. 

\section{Specific elastic symmetries and weak anisotropy}\label{spec}

\subsection{Monoclinic symmetry}

We first provide a general procedure for determining acoustic axes in materials with monoclinic symmetry. 
The off-diagonal elements of $\bQ$ are, for arbitrary anisotropy,  
\begin{eqnarray} \label{od}
Q_{23} &=& c_{56} n_1^2 +c_{24} n_2^2 +c_{34} n_3^2+ (c_{23}+c_{44})n_2n_3+(c_{36}+c_{45}) n_1n_3  + (c_{25}+c_{46})n_1n_2, 
\nonumber \\
Q_{31} &=& c_{15} n_1^2 +c_{46} n_2^2 +c_{35} n_3^2 + (c_{45}+c_{36}) n_2n_3 +(c_{13}+c_{55})  n_1n_3  + (c_{14}+c_{56}) n_1n_2, \quad 
\\
Q_{12} &=& c_{16} n_1^2 +c_{26} n_2^2  +c_{45} n_3^2 + (c_{25}+c_{46}) n_2n_3 +(c_{14}+c_{56})  n_1n_3  + (c_{12}+c_{66}) n_1n_2, 
\nonumber
\end{eqnarray}
Let $n_1$ be normal to a plane of symmetry, so that the moduli $c_{i5}$ and $c_{i6}$, $i=1,2,3,4$, all vanish.  It is assumed that the acoustic axis lies in the plane $n_2=0$, which is possible if we allow the original acoustical tensor to be rotated about $\be_1$ by an as-yet unknown angle, say $\theta_0$. In this case, 
\begin{equation} \label{od2}
Q_{23} = c_{56} n_1^2  +c_{34} n_3^2  ,\quad Q_{31} = 
 (c_{13}+c_{55})  n_1n_3,\quad Q_{12} = 
(c_{14}+c_{56})  n_1  n_3 
\end{equation}
and hence, from eqs. \rf{by}, 
\begin{eqnarray} \label{od3}
&& q_1^2 = ( c_{14}+c_{56} )(c_{13}+c_{55}) \frac{n_1^2  n_3^2}{ c_{56} n_1^2  +c_{34} n_3^2}, 
\\  
&& q_2^2 = \big(\frac{ c_{14}+c_{56} }{c_{13}+c_{55}} \big)(c_{56} n_1^2  +c_{34} n_3^2) , 
\quad  
q_3^2 = \big(\frac{c_{13}+c_{55}}{ c_{14}+c_{56} } \big)(c_{56} n_1^2  +c_{34} n_3^2),
\end{eqnarray}
Using $Q_{22} = c_{66} n_1^2  +c_{44} n_3^2$ and $Q_{33} = c_{55} n_1^2  +c_{33} n_3^2$ the condition  $D_2=D_3$ becomes
\begin{equation} \label{od6}
\bigg[ c_{66} -c_{55} +\bigg( \frac{ c_{14}+c_{56} }{c_{13}+c_{55}}   +  \frac{c_{13}+c_{55}}{ c_{14}+c_{56} } \bigg) c_{56}\bigg]\, n_1^2 =  \bigg[ c_{33}-c_{44} + \bigg( \frac{ c_{14}+c_{56} }{c_{13}+c_{55}}   -  \frac{c_{13}+c_{55}}{ c_{14}+c_{56} } \bigg) c_{34} \bigg]\, n_3^2 .\quad
\end{equation}
Similarly, using $Q_{11} = c_{11} n_1^2  +c_{55} n_3^2$, condition $D_1=D_2$ and the identity $n_1^2+n_3^2=1$, provides a single but lengthy equation involving the moduli only, which must be solved for $\theta_0$.   

\subsection{RTHC crystals}

More explicit and useful methods exist for  materials with symmetries higher than monoclinic.  Thus, Mozhaev et al. \cite{mozhaev2001} derived a method similar to that for the monoclinic material above for trigonal materials, but which gives an explicit expression for $\sin 3 \theta_0$.   All other symmetries above monoclinic, viz. orthorhombic, tetragonal, hexagonal and cubic,  are included in the category of ``RHTC crystals".  These symmetries have been  considered in detail by  Boulanger and Hayes \cite{bh98a,bh98} who showed, using a result of Fedorov and Fedorov \cite{fedfed}, that  acoustic axes in RTHC crystals can be completely classified by relatively simple algebraic means.  

The key to the analysis of Boulanger and Hayes \cite{bh98a,bh98} is the observation of Fedorov and Fedorov that the acoustical tensor for RTHC materials can be expressed in the form of eq. \rf{s7} with $s=+1$ where 
$D_i$, $i=1,2,3$ and the vector $\bN$ have explicit forms: 
\begin{equation} \label{ff2}
D_i = \sum\limits_{j=1}^3 a_{ij} n_j^2\, , \quad  \bN = ( \beta_1 n_1 , \beta_2 n_2 , \beta_3 n_3 ) .  
\end{equation}
The constants $a_{ij}$  and $\beta_i$  can be expressed in terms of the moduli \cite{bh98a,bh98}.  
The results of Boulanger and Hayes are  rederived in Section \ref{gi}  using geometrical,  and perhaps more general,  arguments which shed new light on the structure of the acoustic axes in RTHC crystals.  First we consider the separate but important case of materials with weak anisotropy. 

\subsection{Acoustic axes in weakly anisotropic materials}

A material is defined as weakly anisotropic if the stiffness tensor is of the form ${\bf C} = {\bf C}^{(0)} + \epsilon {\bf C}^{(1)}$, where ${\bf C}^{(0)}$ is isotropic and $|\epsilon|\ll 1$.  Accordingly, the acoustic axis function of eq. \rf{ac8} can be expanded using ${\bf Q}({\bf n}) = {\bf Q}^{(0)}({\bf n}) + \epsilon {\bf Q}^{(1)}({\bf n})$, as 
\begin{equation} \label{ac8a}
 [ \bm,\, \bQ \bm,\, \bQ ^2\bm ]  = 
 f^{(0)}(\bm ) + \epsilon  f^{(1)}(\bm ) + \epsilon^2  f^{(2)}(\bm ) + \epsilon^3  f^{(3)}(\bm ) . 
\end{equation}
The  term $f^{(0)}(\bm )$ is identically zero by virtue of the fact that ${\bf C}^{(0)}$ is isotropic, and hence the condition for acoustic axes becomes
\begin{equation} \label{ac8b}
  f^{(1)}(\bm ) = -  \epsilon  f^{(2)}(\bm ) - \epsilon^2  f^{(3)}(\bm )  .   
\end{equation}
After some simplification, it can be shown that 
\begin{eqnarray} \label{ac8c}
  f^{(1)}(\bm ) &=& -  (\lambda^{(0)} + \mu^{(0)} )^2 \, ( \bn\cdot\bm )\,  [ \bm,\, \bn,\, \bQ^{(1)}\bP \bm ]  
  \nonumber \\
  &=&  ( \bn\cdot\bm )\,  \bm\cdot \bosy{\psi}\bm \, , 
\end{eqnarray}
where $\lambda^{(0)}$ and $\mu^{(0)}$ are the isotropic Lam\'e moduli,  $\bP \equiv \bI - \bn \otimes\bn$, and the  tensor $\bosy{\psi}$ is defined by 
\begin{equation} \label{ac8d}
 \psi_{ij} = - \frac12 (\lambda^{(0)} + \mu^{(0)} )^2 \, ( e_{ikl}n_k S_{jl} +e_{jkl}n_k S_{il}), \quad \mbox{and} \quad  {\bf S} = \bP \bQ^{(1)} \bP    \, . 
\end{equation}
Equations \rf{ac8b} and \rf{ac8c} imply 
\begin{equation} \label{ac8e}
 \phi_{ijk} = \psi_{ij}  n_k  + \psi_{ik}  n_j + \psi_{jk}  n_i + \mbox{O}(\epsilon). 
\end{equation}
Hence, to leading order in $\epsilon$, the condition for the existence of acoustic axes reduces to the requirement that the {\it second order} tensor $\bosy{\psi}$ vanish.  We define the {\it weakly anisotropic acoustic axis condition}  as $\bosy{\psi}=0$. 
Some insight into $\bosy{\psi}$ follows from the general properties of $\bosy{\phi}$, from which it may be shown that $\bosy{\psi}\bn = 0$, tr $\bosy{\psi}=0$ and $\bosy{\psi}^2 =   \bP$ tr$(\bosy{\psi}^2)/2$.   Furthermore, it can be shown that $\bosy{\psi}$ vanishes iff ${\bf S} = \lambda \bP$, where  $\lambda = \frac12$tr${\bf S} $.  Since ${\bf S}$ is rank degenerate by virtue of its definition, $\bosy{\psi} $ vanishes iff the two non-zero eigenvalues of ${\bf S}$ are equal.  Let $\lambda_1$ and $\lambda_2$ be the eigenvalues, then the  
weakly anisotropic acoustic axis condition  $2(\lambda_1^2+\lambda_2^2)-(\lambda_1+\lambda_2)^2=0$ can be expressed, using the identities $\mbox{tr}{\bf S}^n=\lambda_1^n+\lambda_2^n$, as
\begin{equation} \label{ac8f}
 2\mbox{tr}{\bf S}^2 -  (\mbox{tr}{\bf S})^2= 0. 
\end{equation}
 
Equation \rf{ac8f} is interesting in that it apparently provides a single condition for the existence of acoustic axes.  However, its simplicity begs the question of whether satisfaction of this weakly anisotropic acoustic axis condition is equivalent to the exact condition \rf{ca1}.   In order to answer this we assume, with no loss in generality, that the isotropic medium is chosen such that $\rho c_L^2 = \bn \cdot \bQ\bn$ and 
$\rho c_T^2 = \frac12$tr$ \bP \bQ\bP$, where $c_L = (\lambda^{(0)} + 2\mu^{(0)})^{1/2}/\rho^{1/2}$ and $c_T = (\mu^{(0)} /\rho)^{1/2}$ are the longitudinal and transverse wave speeds, respectively.  This implies that $\mbox{tr}{\bf S}=0$ automatically.  The weakly anisotropic acoustic axis condition  becomes ${\bf S}=0$, and when it is satisfied, it can be shown that the acoustic tensor must be of the form $\bQ =  \bQ^{(0)} + \epsilon (\bp \otimes \bn +\bn \otimes \bp )$ for some $|\bp|=1$, $\bp\cdot\bn =0$.  Thus, $\rho c_T^2$ is an exact eigenvalue with transverse polarization in the direction $\bp \wedge \bn$.   A simple analysis shows that the quasi-transverse eigenvalue is
\begin{equation} \label{qt}
\rho c^2 = \rho c_T^2 - \epsilon^2 \big( \rho c_L^2- \rho c_T^2\big)^{-1} + \mbox{O}( \epsilon^4). 
\end{equation}

Hence, when the weakly anisotropic acoustic axis condition is satisfied, one of the split transverse waves is a pure transverse wave with speed $c_T$, and the other has speed $c_T +\mbox{O}( \epsilon^2)$.  The direction $\bn$ is an acoustic axis iff these speeds are equal, and occurs if $\epsilon \equiv 0$, in which case all directions perpendicular to $\bn$ are transverse directions and a pure longitudinal wave exists in direction $\bn$.   Thus, we conclude that the weakly anisotropic acoustic axis condition predicts an acoustic axis iff $\bn$ is also a longitudinal axis. This additional requirement greatly restricts the utility of the weakly anisotropic test for acoustic axes, to the point that it is virtually useless in practice.   In other words, if one calculates slowness surfaces to leading order in the weak anisotropy, the occurrence of double roots is not in general an indication of real acoustic axes but coincides with acoustic axes only if $\bn$ is also a longitudinal axis.

\section{Geometrical interpretation of the acoustic axes conditions}\label{gi}

The analysis in Section \ref{secaa} motivated the previously known  form of the conditions for the existence of acoustic axes \cite{khatkevich1962}.  Having found the more general form involving the acoustic axis tensor in Sections \ref{cis} and \ref{aat}, we now return to the ``{\it ad hoc}" analysis based on the assumed form \rf{form1} and show that it also explains the variety of possibilities contained in the third order tensor $\bosy{\phi}$. 

Without any loss in generality, we assume that $D_1 > D_2 > D_3$.  Then by expressing $\bD$ in  biaxial form similar to eq. \rf{s1}, the acoustic tensor  can be written  
\begin{equation} \label{ff3b}
\bQ (\bn) = D_2 \, \bI  + \frac12 \big( \bd_+ \otimes \bd_- + \bd_- \otimes \bd_+ \big) +\bN \otimes \bN , 
\end{equation}
where 
\begin{equation} \label{ff3}
\bd_\pm = D_{12}^{1/2}\, \be_1 \pm D_{23}^{1/2}\, \be_3, 
\end{equation} 
and the abbreviated notation 
\begin{equation} \label{abb}
D_{ij} \equiv  D_i - D_j,
\end{equation}
 is used.    
Thus,   
 $\bQ $ is uniaxial when the sum of terms in eq. \rf{ff3b}  reduces to the standard form for the  uniaxial tensor, eq. \rf{s7}.   The question can then be phrased as follows: assuming $D_1 > D_2 > D_3$, what is the constraint upon $\bN$ that ensures $\bQ $ of eq. \rf{ff3b} is uniaxial?     
The answer requires consideration of various possibilities for $\bN$, which we now examine in detail. 

First, noting that $D_1$ is the largest of the $D_i$, it is clear that there is no way that $\bQ$ can be uniaxial if  $\bN \parallel \be_1$.  If $\bN$ is in the $\be_2$ ($\be_3$) direction, then its magnitude must be exactly $D_{12}^{1/2}$ ($D_{23}^{1/2}$) in order for $\bQ$ to be oblate (prolate).   However, it is not sufficient that $\bN$ lie along one of the principal directions of $\bD$, since it must have a very specific magnitude if $\bQ $ is to be uniaxial.  For instance,  $\bN \parallel \be_2$ and $|\bN| = D_{12}^{1/2}$ implies $Q_{11}-Q_{22}=0$ with the addition constraint  $\bN \parallel \be_2$, which cannot hold in general, as  illustrated by  the specific example of RTHC crystals, see \cite{bh98}. 
The cases of  $\bN$ in coordinate planes and out of coordinate planes  must therefore be considered separately. 
If we assume that  $\bN$ lies in the plane spanned by $\be_1$ and $\be_2$,   
then in order that $\bQ$ is uniaxial we must have   
\begin{equation} \label{re}
D_1\be_1\otimes \be_1 + D_2 \be_2\otimes \be_2  +\bN \otimes \bN  = D' \, (\be_1\otimes \be_1 + \be_2\otimes \be_2),
\end{equation} 
 for some $D'$.  However,  writing
\begin{eqnarray} \label{how}
&&D_1\be_1\otimes \be_1 + D_2 \be_2\otimes \be_2 
\nonumber \\
&&\, \qquad \, = \frac12 (D_1+D_2) \big(\be_1\otimes \be_1 + \be_2\otimes \be_2\big)
+ \frac12 (D_1-D_2) \big(\be_1\otimes \be_1 - \be_2\otimes \be_2\big),
\end{eqnarray}
then it it clear from the second term in the right member, specifically the combination $(\be_1\otimes \be_1 - \be_2\otimes \be_2)$, that $\bN = q_1\be_1 + q_2\be_2$ cannot satisfy eq. \rf{re} unless $q_1= 0$ and $q_2^2 = D_1-D_2$.  Thus, we again find that the acoustical tensor becomes uniaxial only if $\bN$ is along a coordinate axes, and then only if a  specific relation holds, which cannot hold in general.  A similar negative result is found if  $\bN$ is sought in the plane spanned by $\be_2$ and $\be_3$:  the only possible $\bN$ that leads to uniaxility is $\bN = \pm D_{23}^{1/2}\, \be_3$.  
 
We are thus led to consider $\bN$  in the plane of $\be_1$ and $\be_3$, or equivalently, the plane spanned by $\bd_+$ and $\bd_-$.  Assume the form 
\begin{equation} \label{ff4}
\bN = \frac12 ( \alpha_+ \bd_+ - \alpha_- \bd_- ), 
\end{equation}
for some constants $\alpha_\pm$, which implies, using eq. \rf{ff3b}, that  
\begin{equation} \label{ff5}
\bQ=  D_2 \, \bI  + \frac14 \bigg[ \alpha_+^2 \bd_+\otimes \bd_+ + \alpha_-^2 \bd_-\otimes \bd_-
+ \big( \bd_+ \otimes \bd_- + \bd_- \otimes \bd_+ \big) ( 2 - \alpha_+\alpha_-)\bigg] \, . 
\end{equation}
This reduces to the uniaxial form \rf{s7} iff $\alpha_+\alpha_- = 1$,
and then  
$ \bQ (\bn)  =  D_2 \, \bI  + \bu \otimes \bu $, 
where $\bu  =  ( \alpha_+ \bd_+ + \alpha_- \bd_- )/2$. 
Thus, at uniaxiality, 
\begin{eqnarray} \label{ff8b}
\bN &=& \frac12 \big[ ( \alpha_+  - \alpha_- )D_{12}^{1/2}\, \be_1  + ( \alpha_+  + \alpha_- )D_{23}^{1/2}\, \be_3 \big], 
\\
\bu &=&  \frac12 \big[ ( \alpha_+  + \alpha_- )D_{12}^{1/2}\, \be_1  + ( \alpha_+  - \alpha_- )D_{23}^{1/2}\, \be_3 \big]. 
\end{eqnarray}
Removing the dependence upon the parameter $\alpha_+$ (and $\alpha_- = 1/ \alpha_+$) implies that $\bN$ and $\bu$ lie on complementary hyperbolae in the $\be_1 , \be_3$ plane: 
\begin{equation} \label{ff9}
\frac{q_3^2}{D_{23}} - \frac{q_1^2}{D_{12}}=  1, \qquad 
\frac{u_1^2}{D_{12}} - \frac{u_3^2}{D_{23}} = 1. 
\end{equation}
The eigenvalues are $\{D_0, \, D_2, \, D_2\}$, where the largest eigenvalue $D_0$ follows from eq. \rf{ff10} as 
$ D_0 = D_2+u^2 = D_1-D_2+D_3 + q^2 $. 
Noting   
\begin{equation} \label{ff10}
q^2 = \big( \frac{\alpha_+  - \alpha_- }{2} \big)^2 \, D_{13}  + D_{23} \ge D_{23}, \qquad
u^2 = \big( \frac{\alpha_+  - \alpha_- }{2} \big)^2 \, D_{13}  + D_{12} \ge D_{12}, \qquad
\end{equation}
 we see that  
$D_0\ge  D_1$.  Also, 
$\bu =  \big( q_3 D_{12} , \, 0, \, q_1 D_{23} \big)\, ( D_{12}D_{23})^{-1/2}$,
indicating  that the plane of circular polarization is spanned by $\be_2$ and $ \big( q_1 D_{23}  , \, 0, \, -q_3 D_{12} \big)$. 

Boulanger and Hayes \cite{bh98} obtained the  condition $\rf{ff9}_1$  for RTHC crystals; specifically, it is identical to  eq. (3.6) of \cite{bh98}.  This is also equivalent to $\big( Q_{11} - Q_{22}\big) \big( Q_{33} - Q_{22}\big)-   Q_{31}^2 = 0$, which in turn is equivalent to  $\phi_{112}=0$ or $\phi_{332}=0$.   Equation $\rf{ff9}_1$ is an instance of the specific application of the condition $\phi_{iij}$ (no sum), as compared with the original derived requirements \rf{w6}. 

We now consider $\bN$ out of the coordinate planes, so that no component is zero.  The eigenvalues of $\bQ $ are the zeroes of $g(\lambda)$ = det$(\bD + \bN\otimes\bN - \lambda \bI)$.  This cubic in $\lambda$  takes on the values $g(D_i) = q_i^2D_{ij}D_{ik}$, where $\{i,j,k\}$ is any permutation of $\{1,2,3\}$. Thus, $g(D_1)>0$, $g(D_2)<0$ and $g(D_3)>0$, indicating that the eigenvalues of $\bQ$ must satisfy $D_3< \lambda_3 < D_2< \lambda_2 < D_1< \lambda_1$.   In particular, the eigenvalues are distinct as long as $q_1q_2q_3\ne 0$ and $D_1> D_2 > D_3$ (see Section 4 of \cite{bh98a}).   Therefore,  the latter constraint  must be relaxed and either $D_{12}=0$ or $D_{23}=0$ assumed, that is 
\begin{equation} \label{that}
\bQ = D_2\bI +D_{12} \, \be_1\otimes\be_1 + \bN\otimes\bN  
\quad \mbox{or} \quad 
\bQ = D_2\bI -D_{23} \, \be_3\otimes\be_3 + \bN\otimes\bN , \quad 
\end{equation}
respectively.  The only way $\rf{that}_1$ can be uniaxial is if $q_1=0$ and $q^2 = D_{12}$, implying the oblate 
$\bQ = D_1\bI -D_{12} \, \be\otimes\be$ where $\be = \be_1\wedge \bN/|\bN |$.  Similarly, in order for $\rf{that}_2$ to be uniaxial, $\bN$ must be aligned with $\be_3$.   Both solutions again place $\bN$ in coordinate planes, and therefore the only possibility for uniaxiality with $\bN$ out of the coordinate planes is that $\bD$ is spherical, that is, $D_1=D_2=D_3$.  This was derived previously in Section \ref{secaa}, and was also found by Boulanger and Hayes for RTHC crystals, see eq. (4.7) of \cite{bh98}.

Finally, we consider the possibility $\bq = 0$, in which case $\bQ$ is diagonal and uniaxiality occurs when two of the diagonal elements are equal.  As noted in Section \ref{aat}, the only nonzero element of $\bosy{\phi}$ is $\phi_{123}$  when  $\bQ$ is diagonal. Hence, this particular case of $\bq$ for acoustic axes corresponds to the unique condition $\phi_{123}=0$. 

\subsection{Discussion}

We have seen that the geometrical derivation  of the acoustic axes conditions reproduces all elements of the acoustic axis tensor $\bosy{\phi}$.  These results  provide   new insights into the structure of symmetric second order tensors, beyond the specific interest in degenerate acoustical tensors.  Consider, for instance, the stress tensor of a body in a state of triaxial stress, for which the principal stresses are distinct, $\sigma_1> \sigma_2>\sigma_3$.  The body is then subject to an additional {\em uniaxial} stress of the form $T \bn\otimes \bn$, which is tensile or compressive depending as $T$ is positive or negative, respectively.   Suppose that the values of $\sigma_1,\, \sigma_2,\, \sigma_3$ and $T$ are all fixed (given), and the question is posed: what choice or choices of the direction of uniaxial additional stress will result in a state of total stress that is uniaxial?   The answer follows immediately from the acoustic axes analysis: the direction $\bn$ must lie in the plane of the least and greatest stress, i.e.  $\bn = n_1\be_1 + n_3\be_3$, where, by analogy with eq. $\rf{ff9}_1$, 
\begin{equation} \label{str}
\frac{n_3^2}{\sigma_2-\sigma_3} - \frac{n_1^2}{\sigma_1-\sigma_2} = \frac{1}{T}.   
\end{equation}
The state of uniaxial stress can only be achieved if 
$T> \sigma_2-\sigma_3\quad $ or $ T < -(\sigma_1-\sigma_2)$, and 
the resultant stress is 
\begin{equation} \label{stres}
\bosy{\sigma} = \sigma_2 \bI  +(\sigma_0-\sigma_2)\, \be\otimes\be\, , 
\end{equation}
where the third stress $\sigma_0$ and its direction $\be$ are given by
\begin{equation} \label{th}
 \sigma_0= T + \sigma_1-\sigma_2+\sigma_3,  \qquad\be = 
\, 
\frac{ 
 (\sigma_1-\sigma_2) \, n_3\be_1 + (\sigma_2-\sigma_3) \, n_1\be_3 }{ [(\sigma_1-\sigma_2)^2n_3^2+(\sigma_2-\sigma_3)^2n_1^2]^{1/2}}
 \, . 
 \end{equation}
 It follows that the stress is prolate (oblate) with  $\sigma_0 > \sigma_1 $ ($\sigma_0 < \sigma_3$), if the applied uniaxial stress is tensile (compressive).   Note that the stress analog problem is slightly different from the acoustic axes issue since $T$ can be negative as well as positive.  

\section{Conclusion}

Conditions for the existence of acoustic axes can be formulated in terms of a simple geometrical framework, as in Sections \ref{secaa} and \rf{gi}, or in terms of the rigorous requirements on the third order acoustic axis tensor $\bosy{\phi}$ of Section \ref{aat}.  Specifically, $\bosy{\phi}=0$ at acoustic axes, which in turn involves a multiplicity of identities some of which are often automatically satisfied, as in symmetry planes.  The case of RTHC crystals is best understood in terms of the geometrical approach.   

\appendix
\section*{Appendix}

\section{Proof of equation \rf{s1}}
The unit vectors $\bfa_1$ and $\bfa_2$ are determined by finding isotropic bivectors $\bv_1, \bv_2$ that solve  
\begin{equation} \label{s2}
\bv\cdot\bosy{\Lambda}\bv = 0.
\end{equation}
A {\em bivector} \cite{bhbiv} is a complex valued vector,  $\bv = \bp + i \bq$.  The dot product  is defined  as for real vectors (no complex conjugates are involved), thus, $\bv_1 \cdot \bv_2 
= \bp_1 \cdot \bp_2 - \bq_1 \cdot \bq_2 + i ( \bp_1 \cdot \bq_2 + \bp_2 \cdot \bq_1)$.  The bivector $\bv= \bp + i \bq$ is called  {\em isotropic} if $\bv \cdot \bv = 0$, and occurs iff $\bp$ and $\bq$ are orthogonal and of equal magnitude. 
Zeroes of eq. \rf{s2} may be sought in the form $\bv = (z^2 - 1) \be_1 +i(z^2+1)\be_2 + 2z \be_3$, which is clearly an isotropic bivector for all complex-valued $z$.   Equation \rf{s2} is a quadratic  in $\bv$, homogeneous in the components of $\bv$, and becomes a quartic in the complex variable $z$.  The roots occur in pairs, since it is clear that if $\bv$  is a zero, and so also is $\bv^*$, hence $-1/z^*$, which gives the same bivector.  Thus, two distinct bivectors, $\{\bv_1, \, \bv_2\}$,  solve eq. \rf{s2}, although they are not unique since any other pair, $\{\bv_1', \, \bv_2'\} = \{e^{i\psi_1}\bv_1, \, e^{i\psi_2}\bv_2\}$, $\psi_1,\, \psi_2 \in R$, also solves eq. \rf{s2}. The non-uniqueness indicates that   the planes spanned by the bivectors   are of importance, rather than the bivectors {\it per se}.  The real unit vectors $\bfa_1,\ \bfa_2$ are the normals to these planes,  since this choice of directions for $\bfa_1, \ \bfa_2$ ensures that $\bosy{\Lambda}$ of eq. \rf{s1} has $\bv_1, \ \bv_2$ as zeroes in  eq. \rf{s2}.    It may be shown that  $\cosh \gamma_j\ \bfa_j =  \cos \theta_j\ \be_1 + \sin \theta_j\ \be_2 + \sinh \gamma_j\ \be_3$, where $z_j = e^{-(\gamma_j +i\theta_j)}$, $j=1,2$.      
Let  $\bd = \bfa_1\wedge \bfa_2 /|\bfa_1\wedge \bfa_2|$, and referring to eq. \rf{s1}, 
\begin{equation} \label{s3}
\lambda = \bd\cdot \bosy{\Lambda}\bd, \qquad  \lambda^\prime = \big(tr\ \bosy{\Lambda} - 3\bd\cdot \bosy{\Lambda}\bd\big)/\bfa_1\cdot \bfa_2\ .  
\end{equation}
This completes the construction of the general  form of $\bosy{\Lambda}$ in eq. \rf{s1}. 

It is interesting to note that the bivectors $\bv_1$ and $ \bv_2$ satisify 
\begin{equation} \label{intnote}
\bosy{\Lambda}\bv_j = \lambda \bv_j + \mu_j \bfa_j \quad \mbox{no sum}, 
\end{equation}
where $\mu_1 = (\lambda^\prime /2) \bfa_2 \cdot \bv_1$ and $\mu_2 = (\lambda^\prime /2) \bfa_1 \cdot \bv_2$.  With no loss in generality, we may express the bivectors as $\bv_j = e^{i\psi_j} ( \bd + i \bd \wedge \bfa_j) $ (no sum) for some $\psi_1,\, \psi_2 \in R$.  A straightforward calculation gives
\begin{equation} \label{intnote2}
\mu_1 = i  e^{i\psi_1} (\lambda^\prime /2) |\bfa_1\wedge \bfa_2| , 
\qquad
\mu_2 = -i  e^{i\psi_2} (\lambda^\prime /2) |\bfa_1\wedge \bfa_2|,  
\end{equation}
indicating that $| \mu_1 | = | \mu_2 |$.  If $\psi_1 =  \psi_2 \pm \pi \equiv \psi $ then $ \mu_1  =  \mu_2 \equiv \mu$ and we have the general result 
\begin{equation} \label{intnote3}
\bosy{\Lambda}\bv_j = \lambda \bv_j + \mu \bfa_j \quad \mbox{where}\quad 
\mu = i  e^{i\psi} (\lambda^\prime /2) |\bfa_1\wedge \bfa_2| . 
\end{equation}
In particular, the bivectors are eigenvectors with the same eigenvalue iff the bivectors are coplanar, i.e. iff $\bfa_1\wedge \bfa_2 = 0$.  This reproduces the result of Boulanger and Hayes \cite{bhbiv} that $\bosy{\Lambda}$ is uniaxial iff the the bivectors are also eigenvectors.

\section{Cartan decomposition of the acoustic axis tensor}

A  procedure is  described for determining the Cartan decomposition of the tensor $\bosy{\phi}$.  The method, which can be generalized to tensors of other order, is based on finding the irreducible representation of the elements of $\bosy{\phi}$ under the action of SO(2). 
Let $\be_3$ be the preferred direction for rotation under SO(2), and write $\bm = m_\alpha\be_\alpha + m_3\be_3$, where $\alpha = 1,2$.  Thus, 
\begin{equation} \label{ap1}
f(\bm)= \phi_{\alpha\beta\gamma}\, m_\alpha m_\beta m_\gamma + 3\phi_{\alpha\beta 3}\, m_\alpha m_\beta m_3 
+ 3\phi_{\alpha 3 3}\, m_\alpha m_3^2 + \phi_{333}\, m_3^3,
\end{equation}
where  the fact that $\bosy{\phi}$ is totally symmetric has been used.  Introduce the angle $\theta$ via $m_1 = r \cos\theta$, $m_2 = r \sin\theta$, where $r^2 = 1-m_3^2$, then 
\begin{eqnarray} \label{ap2}
\phi_{\alpha\beta\gamma}\, m_\alpha m_\beta m_\gamma 
&=& \big( \phi_{111} \cos^3\theta +\phi_{222} \sin^3\theta + 3\phi_{112} \cos^2\theta\sin\theta
+ 3\phi_{122} \cos\theta\sin^2\theta \big) \, r^3, \qquad 
\nonumber \\
3 \phi_{\alpha\beta 3}\, m_\alpha m_\beta m_3 &=& 
\big(  \phi_{113}\cos^2\theta +\phi_{223}\sin^2\theta +2 \phi_{123}\cos\theta \sin\theta \big)\, 3r^2m_3
\\
3\phi_{\alpha 3 3}\, m_\alpha m_3^2 &=& 
\big(  \phi_{133}\cos\theta +\phi_{233}\sin\theta )\, 3rm_3^2\, .
\nonumber
\end{eqnarray}
Applying simple trigonometric relations, implies 
\begin{eqnarray} \label{ap3}
f(\bm)&=& \frac14(\phi_{111} -3\phi_{122})r^3 \cos 3\theta + \frac14(-\phi_{222} +3\phi_{112})r^3 \sin 3\theta
\nonumber \\
&& +(\phi_{113}-\phi_{223})\frac32 r^2 m_3 \cos 2\theta + \phi_{123}3r^2 m_3 \sin 2\theta
\nonumber \\
&& + \big[ m_3^2 \phi_{133} + \frac{r^2}{4}(\phi_{111}+\phi_{122})\big] 3r \cos \theta 
+ \big[ m_3^2 \phi_{233} + \frac{r^2}{4}(\phi_{222}+\phi_{112})\big] 3r \sin \theta \qquad 
\nonumber \\
&& + (\phi_{113}+\phi_{223})\frac32 r^2 m_3 + \phi_{333} m_3^3\, . 
\end{eqnarray}
Using the harmonic property \rf{phi3}, it can be  deduced by inspection of eq. \rf{ap3} that the quantities
\begin{eqnarray} \label{irr}
&& \phi_{333},
\nonumber \\
&& \phi_{133} \pm i \phi_{233},
\nonumber \\
&& \phi_{113} - \phi_{223} \pm i 2\phi_{123},
\nonumber \\
&& \phi_{111} - 3\phi_{122} \mp i ( \phi_{222} - 3\phi_{112}),
\end{eqnarray}
correspond to the coefficients of the partition of the $7-$dimensional space of third order totally symmetric harmonic tensors into $1+2+2+2$ subspaces as in eq. \rf{irr2}.


\end{document}